# A model for the orbital modulation of PSR J0737-3039B

Zhu-Xing Liang and Yi Liang
18-4-102 Shuixiehuadu, Zhufengdajie, Shijiazhuang, Hebei 050035, China

## Abstract

A model was constructed to explain the orbit modulation of PSR J0737-3039B (referred to as B), which included the following suppositions: (1) B has arrived at the aged period and generally cannot emit observable light; (2) The PSR J0737-3039A (referred to as A) is dynamic and is encircled by dense charged particles. When A revolves along its orbit, some of the particles are scattered and form a particle cloud band along its orbit; (3) When passing the two intersection points of the orbits, B goes through the cloud band and drives the cloud particles to emit light; (4) Generally, B passes twice through the cloud band every orbit cycle; hence, there are two bright phases along the orbit. (5) It is the complexities of the distribution of the cloud particles and B's emission region that cause the bright phases to drift away from the orbit intersections. We also provide two predictions for the coming performance of B to test our model.

## Introduction

The double-pulsar system PSR J0737-3039A/B was discovered nine years ago ((Burgay et al., 2003; Lyne et al., 2004). This system consists of a 22 ms pulsar (referred to as A) and a 2.77 s pulsar (referred to as B) with an orbit period of 2.4 hours. The discovery of this double-pulsar system provided a unique opportunity to study the magnetosphere of the pulsar and the pulse emission mechanism. The most intriguing behaviors of B are the dramatic orbital modulation of emission and the disappearance of its total radio emissions in March 2008 (Perera et al. 2010). These strange behaviors imply that the emission mechanism of a pulsar is more complex than previously understood. Some models have been proposed to explain B's orbital modulation (Jenet & Ransom 2004; Zhang & Loeb 2004; Lyutikov 2004, 2005, 2010; Perera et al. 2010). Although the models vary and each has its strong point, all include a common viewpoint: It is A's energy flux that directly impacts B's magnetosphere and causes the orbital modulation. After analyzing all the models, we attempted to put forward a new model to broaden the discussion on B's orbit modulation mechanism.

## Main basis of the model

Our model can be roughly described by the following suppositions:
1. Pulsar B has arrived at its aged period. The reason for the decline of B is that many electrons, which can be driven to emit light, have escaped from B's emission region. The motive power that causes the electrons to escape may be the electric field acceleration. Driven by a very strong electric field, some electrons can emit photons and at the same time escape from the height of the emission region to a higher level. Once B arrives at the aged period, its emission region is short of electrons, and thus the emission is so weak that generally we cannot see it.
2. If some extrinsic electrons come into B's emission region where the magnetic field can drive them to emit photons, the aged B will be active and emit light again.
3. A is a dynamic pulsar encircled by a dense cloud of charged particles similar to the Crab pulsar shown in Fig. 1. When it moves along its elliptical orbit, some of the charged particles are scattered and form a cloud band along its orbit.
4. The orbits of A and B intersect at two intersection points. When B crosses A's orbit and pierces through its cloud band, some of charged particles of the cloud band are driven to emit light by B's magnetic field. Generally, B passes through the cloud band twice every orbit cycle, hence there are two bright phases in its orbit.
5. It is the complexities of both the distribution of cloud particles and B's emission region that cause the bright phases to drift away from the two orbit intersections.

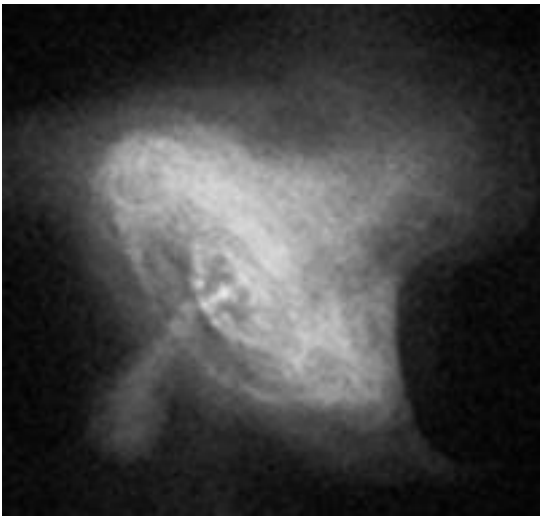

Figure 1. The Crab pulsar (Weisskopf et al., 2000). We suppose that when such a pulsar revolves about another star, some charged particles will be scattered and then form a ring particle cloud along its orbit.

## Manifestations and causes of the bright phases

For convenience of expression, we need to define three terms:
The term "emission region" means the particular region of the magnetic field of a pulsar (here pulsar B). In the emission region, the charged particles can emit light. For the lighthouse model, the emission regions are two hollow-cone shaped spaces formed by the sweeping of two emission cones as the star spins. In particular, if the magnetic axis is orthogonal with the spin axis, the emission region is a discoid space and symmetrical to the equatorial plane. However, under the Magnetic Field Oscillation Model (Liang & Liang 2007), the emission region is always discoid space and symmetrical to the equatorial plane. Irrespective of the model, the configurations of the emission region are slice shaped. This characteristic is very important in the discussion below.

The term "emission band" means a band space. When B orbits, its emission region sweeps a band space along its orbit. Only the charged particles which come to this emission band have the opportunity to emit light.

The term "intersection space" means the space owned by both B's emission band and A's cloud band. Since the two orbits cross each other twice every orbit cycle, the intersection space generally consists of two segments. The sizes of intersection space are close related to the shapes of the two bands and the intersection angle between both bands. The smaller the intersection angle, the smaller the intersection space.

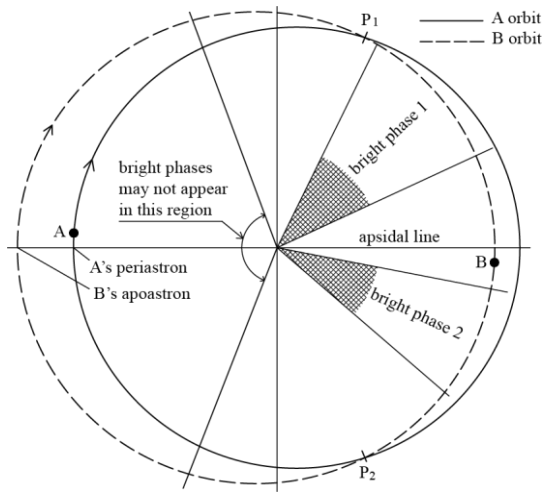

Figure 2. The relationship of the apsidal line, orbit intersections, and bright phases. We predict that the bright phases may not appear at the far left hand side.

According to the main stem of the above model, orbit intersections $P_1$ and $P_2$ in Fig. 2 ought to be included in the two bright phases. However, they are excluded from the bright phases. This result may be caused by manifold affecting factors:

1. The orientation of A's spin axis can affect the density distribution and shape of the cloud band. When A revolves along its orbit, the angle between its equator plane and the direction of its revolution velocity changes continually. If the equator plane is perpendicular to the revolution velocity, the cross-section of the cloud band will be round and stronger. If the equator plane is parallel to the revolution velocity, the cross-section of the cloud band will be strip-shaped and thinner. Hence, the cloud band is non-uniform and twisted. Generally, the cloud band should be asymmetric about the apsidal line. When B revolves, similar to A's cloud band, B's emission band is also non-uniform and twisted. The orientations of B's spin axis can affect the configuration of the emission band. The intersection space is formed by B's emission band and A's cloud band. Therefore, the orientations of the spin axes of A and B can affect the longitude positions of the intersection spaces. The sizes of the intersection spaces and the particle density in the intersection spaces are functions of orbit longitude; hence B's emission flux must be a function of orbit longitude. At any time, as long as B's emission region cannot meet with the cloud band, B's emission will disappear from sight.

2. The so-called bright phases are simply the phases of intersection space. Because all the factors which can affect A's cloud band or B's emission band can also affect their intersection spaces, the bright phases must show complex phenomena. The obvious phenomenon is that the bright phases can drift away from the orbit intersection points $P_1$ and $P_2$.

3. Because the configuration of the intersection spaces evolves with time, the two bright phases evolve continually in flux, profile, longitude position and extent. If the intersection spaces are very small or the particles in the intersection space are very thin, B's emission will disappear such as in March 2008 (Perera et al. 2010). In other words, the total disappearance of B's radio emissions does not imply that the intersection space had totally disappeared, only that the intersection space size or the particle quantity there have become very small. The evolution and disappearance of the bright phases are caused by the advance of the periastron and the precessions of the spin axes of A and B. The advance of the periastron and A's axis precession can change the relationship between the apsidal line and the cloud band. The precession of the spin axis of B can change the relationship between the apsidal line and B's emission band. Therefore, both the advance of the periastron and the precessions of the spin axes can make the two bright phases evolve with time. The extreme case of evolution is the total disappearance of B's radio emissions.

4. Furthermore, the gravitational forces should be considered. They may also change the shape of the cloud band from a perfect ellipse. After considering the effect of the gravitational force of both neutron stars, the center line of cloud band cannot remain at A's orbit. It may be pulled by the gravitation of both neutron stars and shift toward the barycenter of the system. Accordingly, the bright phases may be caused to drift away from intersection points P1 and P2.

5. In addition to the above two bright phases, Pulsar B often shows weak emission at other orbital phases (Lyne et al. 2004). Apart from the cloud band discussed above, there should be other thinner particle clouds along B's orbit. If the sensitivity and signal-noise ratio of observation is greatly improved, the emission can be detected over the entire orbit. As long as the pulsar B goes through the particle cloud, it can emit light. But whether the light can be detected depends on the particle quantity in B's emission regions.

6. The rate of advance of the periastron is about $17°\ yr^{-1}$ (Lyne et al., 2004). However, Perera et al. (2010) reported that the centers of the two bright phases move to higher longitudes at rates of $0.85°$ and $3.1°\ yr^{-1}$ respectively. Both the bright phases move slower than the advance of the periastron. This phenomenon may be caused by the difference between the period of advance of the periastron and the periods of precession for A and B. If the precession periods of both spin axes are longer than the period of the advance of periastron, the evolution tendency of the bright phases should be similar to the process illustrated in Fig. 3.

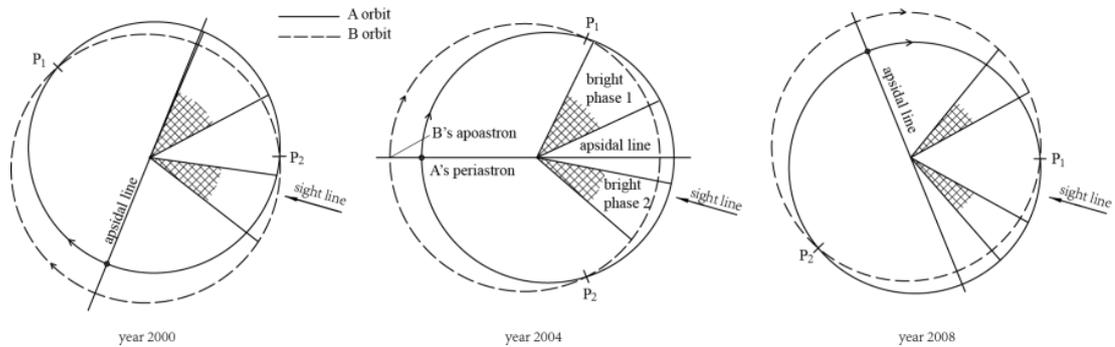

Figure 3. The evolution law of the bright phases. The left case (corresponding to the year 2000) is derived by backward reckoning using the other two cases. Both the apsidal line and the bright phases rotate clockwise, but the apsidal line rotates faster than the bright phases.

## Two predictions

Compared with other models, the distinct characteristic of our model is that it has no direct relationship with A's energy flux. We think A's energy flux can modulate B's emission to form the modulation signal of 22 ms (McLaughlin et al. 2004) but cannot result in B's orbit modulation at the same time.

The discussion above is purely a qualitative analysis and superficial outline. A quantitative analysis needs abundant data and long-term observation. For this aim, at least several cycles of the advance of the periastron are required.

The strength of our model is that it has two testable predictions:
1. The bright phases can appear only on the right hand side of Fig. 2 and cannot appear near B's apoastron (left of Fig. 2). Because the distance between B's emission region and A's cloud band is too great the intersection space cannot appear there.
2. The evolution law is that the bright phases always appear in the vicinity of $P_2$ as in the left panel of Fig. 3, then revolve anticlockwise relative to the apsidal line, and finally disappear in the vicinity of $P_1$ as in the right panel of Fig. 3.

These two characters can be used to test the correctness of this model. Providing the bright phases appear near B's apoastron, it is clear that our model will be ruled out.

## References

Burgay, M. et al. 2003, Nature, 426, 531
Jenet, F. A., & Ransom, S. M. 2004, Nature, 428, 919

Liang, Z. X. & Liang, Y. 2007, arXiv: 0709.4315
Lyne, A. G., et al. 2004, Science, 303, 1153
Lyutikov, M. 2004, MNRAS, 353, 1095
Lyutikov , M. 2005, MNRAS, 362, 1078
Lyutikov ,M. 2010, NewAR, 54, 158
McLaughlin, M. A., et al. 2004, ApJ,613 L, 57
Perera, B. B. P., et al. 2010, ApJ, 721, 1193
Weisskopf, M. C., et al. 2000, ApJ.., 536 L, 81
Zhang, B., & Loeb, A. 2004, ApJ, 614L, 53